
\input harvmac

\def \tt {{\tilde \t}}
 \def \k1 {{1\over
k}}  \def \ov { \over }
  
\def \L {\Lambda}

\def \ra {\rightarrow}

\def \a {\alpha}

\def \sh {{\rm sinh \ }}
\def \Tr {{\ \rm Tr \ }}

\def \ln {{\rm \ ln \  }}

\def \l {\lambda}
\def \p {\phi}

\def \m {\mu }
\def \n {\nu}

\def\g {\gamma}

\def \k {\kappa }
\def \d {\delta}

\def \s {\sigma}
\def \t {\theta}

\def \fourth {{\textstyle{1\over 4}}}

\def \e#1 {{{\rm e}^{#1}}}
\def \const {{\rm const }}

\def \eq#1 {\eqno {(#1)}}
\def \sm {$\s$-model\ }

\def \ov {\over }


\def \p {\phi}

\def \s {\sigma}

\def \d {\delta}
\def \l {\lambda}
\def \m {\mu}
\def \g {\gamma}
\def \n {\nu}

\def \fourth {{1\over 4}}

\def \e#1 {{{\rm e}^{#1}}}
\def \const {{\rm const }}

\def \vp {\varphi}

\def \tg {{\tilde g}}

\def \m {\mu}

\def \ra {\rightarrow}

\def \dvps {{\dot \varphi}^2}

\def \const {{\rm const} }

\def \L {\Lambda}
\def \eq#1 {\eqno{(#1)}}
\def \e {\rm e}
\def \ra {\rightarrow }
\def \e#1 {{\rm e}^{#1}}

\def \a {\alpha}

\def \sh {\ {\rm sinh} \ }

\def \ln { {\rm ln } }
\def \sin {\ {\rm sin} \ }
\def \cos {\ {\rm cos} \ }
\def \tg {\ {\rm tg} \ }
\def \l {\lambda}
\def \p {\phi}
\def \vp {\varphi}
\def  \g {\gamma}

\def \k {\xi}
\def \d {\delta}
\def \s {\sigma}
\def \t {\theta }

\def \pa {\partial}

\def \sqG {\sqrt {-G}}

\def \dl {\dot \lambda}
\def \ddl {\ddot \lambda}
\def \dvp {\dot \varphi}
\def \ddvp {\ddot \varphi}
\def \tt {{\tilde \theta }}

\def\np {  Nucl. Phys. }
\def \pl { Phys. Lett. }
\def \mpl { Mod. Phys. Lett. }
\def \prl { Phys. Rev. Lett. }
\def \pr  { Phys. Rev. }

\baselineskip8pt
\Title{\vbox
{\baselineskip 10pt{\hbox{ }}{\hbox{  }}{\hbox
{Imperial/TP/93-94/36 }
}{\hbox{hep-th/9404191 }} } }
{\vbox{\centerline { On `rolling moduli' solutions
 }\vskip2pt
 \centerline {in string cosmology  }\vskip2pt
 \centerline{
 }
}}
\vskip -5 true pt
\medskip
\centerline{   A.A. Tseytlin\footnote{$^{*}$}{e-mail:
tseytlin@ic.ac.uk}\footnote{$^{\star}$}{\baselineskip8pt
On leave  from Lebedev Physics
Institute, Moscow, Russia.
} }
\smallskip
\smallskip
\centerline {\it Theoretical Physics Group, Blackett Laboratory }
\centerline {\it Imperial College,  London SW7 2BZ, U.K. }
\bigskip\bigskip\bigskip

\centerline {\bf Abstract}
\smallskip
\baselineskip6pt
\noindent
Given a static string solution  with some free constant parameters (`moduli')
it may be possible to construct a time-dependent solution
by just replacing the moduli by some functions of time.
We present   several examples  when such `rolling moduli' ansatz is consistent.
In particular, the anisotropic  $D=4$ cosmological  solution
of Nappi and Witten  can be reinterpreted  as a
time-dependent generalisation of the (analytic continuation of)
$D=3$ `charged black string'  background
with the `charge'  changing with time.
We find some new $D=4$ cosmological solutions
which  are `two rolling moduli' generalisations of the previously known ones.
We also comment on interplay between   duality transformations and
the replacement of moduli by functions of time.

\smallskip
\bigskip

\Date { {April 1994}}
\noblackbox

\baselineskip 20pt plus 2pt minus 2pt

\lref \green { P. Aspinwall, B. Greene and D. Morrison, \pl B303(1993)249; \np
B416 (1994) 414;
E. Wiiten, \np B403(1993)185; preprint IASSNS-93/36. }

\lref \muell   { M. Mueller, \np { B337}  (1990) 37  }
\lref \vene   { G. Veneziano, \pl { B265} (1991) 287.
 }
\lref \gasven  { K.A. Meissner and G. Veneziano, \pl { B267} (1991) 33; \mpl A6
(1991) 3397.
}
\lref\gsvy  {B.R. Greene, A. Shapere, C. Vafa and S.-T. Yau, \np B337 91990) 1
 }
\lref \duff { }
\lref \ts {A.A. Tseytlin, {\it Int. J. Mod. Phys.} { D1} (1992) 223. }
\lref \tsva { A.A. Tseytlin and C. Vafa, \np { B372  } (1992)  443.
 }
\lref \gavem{M. Gasperini, J. Maharana and G. Veneziano, \pl B296 (1992) 51. }
\lref \ishi {N. Ishibashi, M. Li  and A. R. Steif,
         \prl { 67} (1991) 3336.}
\lref \gpr {A. Giveon,
 M. Porrati and E. Rabinovici,  preprint  RI-1-94, hep-th/9401139.}

\lref \sfts { K. Sfetsos and A.A. Tseytlin, \pr { D49} (1994) 2933.}

\lref \gibb {G.W. Gibbons and P.K. Townsend, \np { B282} (1987) 610.
 }
\lref \givkir { A. Giveon and E. Kiritsis, \np B411 (1994) 487. }

\lref \kolu   { C. Kounnas and D. L\" ust, \pl {B289} (1992) 56. }
\lref\givpas  {A. Giveon and A. Pasquinucci, \pl { B294 }(1992) 162.
}
\lref \horne {J.B. Horne and G.T. Horowitz, \np { B368} (1992) 444.}
\lref \cvkut {M. Cveti\v c and D. Kutasov, \pl B240 (1990) 61.  }
\lref \napwi { C. Nappi and E. Witten, \pl { B293 }(1992) 309.
  }
\lref \giv  { A. Giveon, Racah preprint  RI-153-93, hep-th/9310016.}

\lref \ginq{  P. Ginsparg and F. Quevedo, \np { B385 }(1992) 527.}

\lref\kk { E. Kiritsis  and C. Kounnas, preprint CERN-TH.7219/94,
hep-th/9404092.  }

\lref \bs { I. Bars and K. Sfetsos, \mpl  { A7} (1992) 1091;
 E.S. Fradkin and V.Ya. Linetsky, \pl  {B277} (1992) 73;
A.H. Chamseddine, \pl B275 (1992) 63. }


\newsec{Introduction}

Given a static string solution with some free parameters  $\l_i$
  (moduli of the corresponding $2d$ conformal field theory)one may try to look
for cosmological solutions
in which the spatial part remains the same (i.e. the same CFT at each fixed
$t$)
and only the  moduli $\l_i$ change with time.
 A simple
 example is provided by the solution  of \muell\ (see also
\vene\gasven\ginq)  where the moduli  of
 a toroidal  background  and dilaton  evolve with time.
However,  in general
one is not guaranteed to get a solution  by  just replacing $\l_i$ in  an
$N$-dimensional
static solution by some functions of $t$:
 a deformation just along   moduli directions
need not  give   a solution of the full set of string equations in
$N+1$-dimensions
(for  related   remarks  see, e.g., \gibb\gsvy).
Depending on a `spatial'  CFT  the only solution that may result
 may be the  trivial one
only ($\l_i=\const$ and the dilaton linear in time).

The question under which conditions  a  particular trajectory in  a moduli
 space can be identified  with a  string cosmological solution
is of interest in connection with a possibility of a topology change
in  a  process of cosmological evolution.
A topology change  can take place  as  one
moves along  moduli space  of  some  Calabi-Yau  manifolds \green.
It is not known  whether the corresponding
motion  can actually be realised  as a cosmological solution.
A  possibility of a topology change in string cosmology
was recently  discussed  on a simple model in \kk.\foot{The
model discussed in \kk\
is essentially equivalent to the cosmological solution of \napwi\ (based on the
coset  $[SL(2,R)\times SU(2)]/R\times R$)
which is an $O(2,2)$ duality rotation \givpas\  of the   anisotropic
`direct product' cosmological model
of \kolu. The
spatial
part of it  represented by the coset model  $[SU(2)\times U(1)]/U(1)$
 is equivalent to  an  analytic continuation of the `charged black
string'  solution of \horne\ and  was  discussed in detail in connection
with a possibility of
 topology change in \givkir. The topology here changes at the boundary of the
moduli space  and thus may not be of direct interest  (with regions of
different topology being separated by an infinite distance in theory space
\cvkut).
A  smooth topology change  was claimed to happen inside the moduli space
\givkir\giv\
  for  a  \sm deformation of $SU(2)$ WZW model
induced by a particular  $O(2,2)$ duality transformation but  in that case it
is not explicitly known
which  CFT  corresponds to the deformed \sm and  whether the deformation can
occur as a result of cosmological evolution.}

The aim of  this paper  is to make some  simple  observations  (Section 2)
about the  general structure
of  the string cosmological equations for the dilaton and moduli
(assuming that the ansatz $\l_i \to \l_i(t)$ is consistent)
and  to
 present a number of examples
in which the replacement of moduli by  functions of time does,  in fact,
lead to string cosmological solutions (Section 3).
In addition to  reproducing  some previously known solutions
 (in particular,   the $D=4$ solution of \napwi\   can  be  obtained
from   the analytic continuation
of the $D=3$ charged black string solution \horne\  by making   the `charge'
time-dependent;   an equivalent observation was made in \kk)
we will find a number of new $D=4$ anisotropic cosmological string solutions
and discuss the  action of    duality transformations.


\newsec{String cosmological equations for the dilaton and moduli}
Our discussion will be limited to the leading-order
effective equations for the metric, antisymmetric tensor and dilaton couplings
of the bosonic string  \sm.
Consider an $N$-dimensional string  background $G_{ab} (x) , \ B_{ab}(x)  , \
\p (x) \ (a,b=1,...,N)$
and  generalise it to a $D=N+1$-dimensional one by making the fields to depend
on $t$.
This corresponds to a particular gauge choice  in $N+1$ dimensions
$$ ds^2 = \hat  G_{\m\n} dx^\m dx^\n= -dt^2 + G_{ab} (t,x) dx^a dx^b \ , \eq{1}
$$ $$
  \hat B_{\m\n} = (B_{ab}
(t,x), 0 ) \ , \ \ \ \p=\p(t,x) \   .  $$
The leading-order string effective equations  which  follow from
$$ S =  \int d^D x \sqG \  \e{-2\p}   \ \{ {D-26\ov 6}    - {\a'\ov 4} [  \hat
R \ + 4 (\pa_\m \p )^2 - {1\ov 12} \hat H_{\m\n\l}^2]
  + ...  \}  \  ,\eq{2} $$
are given by
$$  \bar \beta_{00}^{G (D)}= \ddot \varphi - {1\over4} G^{ac} G^{bd} (\dot
G_{ab} \dot G_{cd } +  \dot B_{ab} \dot B_{cd }) =0\ ,  \eq{3} $$
$$ \bar \beta_{ab}^{G (D)}= \bar\beta_{ab}^{G (N)} + \ha  [ \ddot G_{ab} -
\dot \varphi \dot G_{ab} -
   G^{cd} (\dot G_{ac}\dot G_{bd} - \dot B_{ac}\dot B_{bd})] =0\ , \eq{4} $$
$$ \bar \beta_{ab}^{B (D)}= \bar \beta_{ab}^{B (N)}
  + \ha  ( \ddot B_{ab} -  \dot \varphi \dot B_{ab}  - 2 G^{cd} \dot G_{c[a}
\dot B_{b]d}  )=0 \ ,  \eq{5} $$
$$ \tilde \beta^{\phi (D)}= \bar \beta^{\phi (D)}- {1\over 4} \hat G^{\m\n }
\bar \beta^{G (D)}_{\m\n}= { c -25\ov 6}    +  {3\over2}\a' ( \ddot \varphi -
\dot \varphi^2)=0\ , \eq{6} $$
$$ { c -25} \equiv 6\tilde \beta^{\phi (N)} + {1} =
   {N -25}- {3\over 2}\alpha'[R-{1\over12}H^2-4(\del\phi)^2+4\nabla^2\phi] \ ,
\eq{7} $$
$$ \bar \beta_{0a}^{G (D)}=   \nabla_c(G^{db} \dot G_{ba})  -  {1\over 2} \dot
B_{db} H_a^{\ db}  + 2 \del_a \dot \varphi -
G^{db} \dot G_{ba} \del_d \varphi=0   \ ,  \eq{8} $$
$$ \bar \beta_{0a}^{B (D)}=- G_{ab}\del_d(G^{bc}G^{de}\dot B_{ce})   + 2   \dot
B_{ab}\nabla^b\varphi=0 \ , \eq{9}  $$
where  dot denotes  a time derivative  and we introduced the  shifted dilaton
$$ \varphi \equiv  2\phi - \ln{\sqrt G}\ .   \eq{10} $$
Eqs.(3)--(6) can be derived from the action  which follows from (2)
($G_{00}=-1$ after the variation)
$$ S=  \int dt d^N x \sqrt{-G_{00}}\  \e{-\vp} \ \{ C    -   G^{00}[ - \dvp^2 +
  \fourth  G^{ac} G^{bd} (\dot G_{ab} \dot
G_{cd }
 +  \dot B_{ab} \dot B_{cd }) ] \}    \ , \eq{11} $$
$$ C\equiv   - {2\over 3\a'} ( c  -25)\  . \eq{12} $$
Suppose $G_{ab} (x, \l) , \ B_{ab}(x,\l)  , \ \p (x,\l)$ represent an
$N$-dimensional string  solution (CFT)  depending on some free  constant
parameters
$\l_i \ (i=1,..,r)$ (for clarity  we shall first treat the constant part of the
dilaton separately from the rest of the moduli).  In that case the
$N$-dimensional Weyl anomaly coefficients $ \bar \beta_{ab}^{G (N)},  \bar
\beta_{ab}^{B (N)}$
 in (4),(5) vanish and $\tilde \beta^{\phi (N)}$   in (7) is
equal to a  constant  ($c$ becomes  the  central charge of the spatial CFT)
which does not depend on the moduli $\l_i$.
Replacing $\l_i$  by functions of time one may try to look for solutions
of (3)--(9) such that
$$ G_{ab} (x,t) = G_{ab} (x, \l(t)) \ , \ \   \ B_{ab} (x,t) = G_{ab} (x,
\l(t)) \ , \ \   \ \phi (x,t) = \p_0(t) + \p(x,\l(t))\ . \eq{13}  $$
In general, such solutions will not exist: the $x$- and $t$-dependence
will not decouple, the constraints (8),(9) will not be satisfied, etc.
Let us, however, proceed by assuming that the `rolling moduli' ansatz  (13)
{\it is}
 consistent
for a particular choice of  a spatial CFT.
Then $\l_i$ and $ \p_0(t)$ will be solutions of the
equations that follow from the action
$$ S=  \int  dt \  \sqrt {-G_{00}} \  \e{-\vp} \   \{ C  + G^{00} [-  \dvp^2  +
 \g_{ij}(\l)  \dl^i \dl^j] + ... \}\ ,     \eq{14}  $$
which follows from (11) upon  integration over $x$ (the consistency of the
ansatz implies that  such integration can be performed).
Here
$ \vp (t) $  is the corresponding shifted dilaton (containing also the constant
factor of the $x$-space volume),
 $\g_{ij}$ is the `Zamolodchikov's
metric' of the  `static' conformal field theory  and dots  stand for  higher
order $\a'$-corrections.
For example, in the case of the toroidal background when $(G_{ab}  , \ B_{ab} ,
\ \p )$
are $x$-independent  the integral over $x$ decouples  and (11)  takes the
manifestly $O(N,N)$ duality-invariant form \gasven\
 $$ S=  \int  dt \ \e{-\vp}  \  [ C -  \dvp^2 -  {1\ov 8} \Tr \bigl( {\dot M
}\eta {\dot
M }\eta )  + ...  ]  \ ,  \eq{15}  $$
 $$M =\pmatrix{ G\inv &  -G\inv B \cr   BG\inv & G- B G\inv B  \cr}\ ,
\ \ \ \ \eta = \pmatrix{ 0 &  I \cr   I&  0   \cr}\ .
\eq{16} $$
In general, the  action  (14)  admits an interesting  interpretation as     an
action  of a particle propagating in
 curved  $r+1$ dimensional space $\l^A=(\vp, \l^i)$ which has $Minkowski$
signature.
Written in the form  invariant under reparametrisations of $t$ this action is
$$ S =  \int dt\  e  \   [ \  C   \   +
\  e^{-2}  \g_{AB} (\L)  \dot \L^A \dot \L^B + ... ]   \ ,  \eq{17}  $$
where we have defined
 the einbein  $e$ and  the  metric $\g_{AB}$ on the space  $\L^A=(\vp, \l^i)$
by
$$ e^2 \equiv  - \e{- 2\vp} G_{00} \ ,  \ \ \ \ \
\g_{AB} = \e{ -2\vp}   \pmatrix{ - 1 & 0  \cr    0  &  \g_{ij}   \cr}\ .
\eq{18} $$
 $\sqrt{-C} $  can be identified with the  mass of the particle,
so that the geodesics are time-like, null or  space-like  for negative, zero or
 positive  $C=- {2\over
3\a'} (c -25) $ where $c$ is the central charge of the  spatial CFT.\foot{ A
related discussion of cosmology  in the presence of the  dilaton and
scalar moduli  fields, and, in particular, the
interpretation of the cosmological equations as geodesic motion in the
`augmented' or `extended'
moduli space $(\L^A)$  appeared in \gibb\  and  was also mentioned in  \gsvy.
The  connection  between the value of  a   central charge (or dimension) and
the
time-like, null or  space-like nature of  geodesics in the moduli space
 was noted in \gsvy. }

The dilaton $\vp$  plays the role of a `time-like' coordinate of the extended
moduli space.
For flat $\g_{ij}$  the  metric in (18) is that of the Milne universe with a
flat spatial
section. The  time
reparametrisation invariance can
be fixed by setting  $G_{00}=-1$, i.e.  $e= {\rm e}^{\vp}$.
An  alternative is  to choose the standard particle theory gauge: $e=1$, i.e.
$G_{00}=  -{\rm e}^{2\vp}$.
In the gauge $G_{00}=-1$ we  find the following evolution equations (cf.
\tsva\ts)
$$ C + \dvps -\g_{jk} \dl^j \dl^k = 2 U  \ \ , \eq{19} $$
$$  \ddl^i + \Gamma^i_{jk}(\g)  \dl^j \dl^k  - \dvp \dl^i = -  {\pa U \over \pa
\l^i} \ \ ,
\eq{20} $$
 $$  \ddvp -  \g_{jk} \dl^j \dl^k  = {\pa U \over \pa \vp }  \ \ ,  \eq{21} $$
where for generality we included   a potential by replacing the `mass term' $C$
in (17) by $V(\l) = C - 2U(\vp, \l) . $
It is interesting to observe that the solution for $\vp$ is {\it universal},
i.e. it does not depend on
a particular $\g_{ij}$.   One  can combine (19) and (21) to get
(in what follows  $U=0$)
$$\ddot y + Cy = 0 \  , \ \ \ \ \  y \equiv \e{-\vp} \ \ , \eq{22} $$
with the solution
(for definiteness we  assume that $C\leq 0$ and  $y(0)=0$)
$$ \vp = \vp_0 - {\ln  \sh } 2bt   \ ,\ \  \ \  4b^2 = -C \ . \eq{23} $$
This  suggest a  close relation between dilaton  and time.
The equation (20) or
$$  \ddl^i + \Gamma^i_{jk}(\g)  \dl^j \dl^k  = \dvp \dl^i \
\eq{24} $$
has a geometrical interpretation  as  an  equation  for a   geodesic in the
$\l^i$-space parametrised by a non-affine parameter.
Defining the affine parameter $s$  (and using (23))
  $$  \ddot s = \dot \vp \dot s \ , \ \ \   \ \  s(t) =  \int^t dt' \e{\vp (t')
}
=
 s_0 +  s_1  \ln \tanh bt   \  ,  \eq{25}  $$
   we find that  (24)  takes the canonical form
$$ {d^2{\l^{i}}\ov ds^2}  +  \Gamma^i_{jk}(\g)  {d\l^j\ov ds} {d\l^k\ov ds} =0
\ ,   \eq{26} $$
implying
$$ \g_{jk} {d\l^j\ov ds} {d\l^k\ov ds} = \m^2 =\const  \ . \eq{26} $$
The solution of (25) is subject to a `zero energy'
constraint  which follows from (19):
$$  C + \dvps  - {\dot s^2 \m^2 } = C +  {4b^2\coth^2 2bt  }
    -  \m^2 \big({2s_1 b  \ov \sinh 2bt}\big)^2 =0  \ . \eq{27} $$
This condition is satisfied if
$$  C=- 4b^2\   , \ \ \ \ \  s_1^2 \m^2=1\  .   $$
The dilaton  (23) expressed in terms of the `proper time' $s$   is
$$ \vp = \vp_0'    +  \  \ln \ { \sinh \mu (s_0-s)   }  \ . \eq{28}  $$
In the case  when $\g_{ij} = \const  $  we get  ${d^2{\l^{i}}/ ds^2}=0$, i.e.
(cf. \muell\gasven)
$$\l_i = \l_{0i} +  \m p_i s= \l_{0i}'  + \ p_i \ \ln \tanh bt \ , \ \ \  \ \
\g_{ij} p^i p^j =1  \ . \eq{29} $$
We can also express $\l_i$ directly in terms of the dilaton.
 In a  reparametrisation-invariant theory  time is an ambiguous notion;  we can
use the component
$G_{00}$ of the metric or the dilaton  itself as a  `dynamical' time.
For example, in string theory we may set $x^0 =\tau$ as a  world-sheet gauge.
Here the situation is similar: the (redefined) dilaton plays the role of time
in the `third-quantised' string picture.


\newsec{Examples of `rolling moduli'  solutions}
Let us now consider  less trivial than torus examples
of   conformal backgrounds  for which
the rolling moduli ansatz (13) produces a consistent  $D=4$
cosmological solution.
We shall  start with the following  (leading-order)  $N=3$  string solution
corresponding to the coset $SU(2)/U(1) \times U(1)$
(i.e., the $SU(2)$ analog  of the  neutral black string background \ishi\horne)
$$ ds^2_N =  dx^2 +  a_1^2 {\rm tan}^2 b'x d\t^2 +  a^2_2   d\tt^2 \ , \
\eq{30} $$ $$ \ \ \p= \p_0 - \ln \cos b'x \
,  \ \ \  \tilde \beta^\p = {N-26\ov 6} -  \a'b'^2  \ .    $$
This is a  conformal background  for arbitrary values of  the constants  $a_i,
 \p_0 $.\foot{To satisfy the central charge condition $\tilde \beta^\p=0$
without introducing extra degrees of freedom one needs to  take
  $b'$ to be imaginary.} Replacing these constants by functions of time
 we find that
the full set of equations for an $N+1$-dimensional solution  (3)--(9)
is satisfied if $a_i(t)  , \p_0 (t)$ solve the equations that
follow from the action  (14) or (17)  with $\g_{ij}= \delta_{ij}\  (i,j=1,2) $,
i.e.
 $$ S=  \int  dt \ \e{-\vp}  \big( C -  \dvp^2  +   \dl_1 ^2 +  \dl_2 ^2\big)
\ , \eq{ 31 }  $$
$$ \ a_i \equiv  \e{\l_i}   \ , \ \ \ \  \vp(t) \equiv 2\p_0 (t) - \l_1(t)
-\l_2(t)\ ,
\ \   \  C=  -{2(N-25)\ov 3\a' }  + 4b'^2 \ .  $$
The equations are thus the same  as in  the case of the  flat two-torus
  so the solution is  given by
(29),(28)
$$  \vp(t)  = \vp_0  - \ln \sinh 2 b t\ , \ \    \ \   4b^2 =-C   \ , \eq{32}
$$
$$  \l_i = \l_{i0}  +  p_i \ln \tanh b't \ , \ \ \  \ \      p_1^2 + p_2^2 =1
\ , \eq{33} $$
so that
$$  ds^2_D = -dt^2 +  dx^2 + \  a_{10}^2{\rm tanh}^{2p_1} bt\  {\rm tan}^2 b'x\
 d\t^2\  +  a^2_{20} {\rm tanh}^{2p_2} bt\  d\tt^2 \ , \eq{34} $$ $$
 \ \vp(x,t) = \vp_0 -
 \ln \sinh 2 b t\ -
  \ln \sin 2 b'x \ .    $$
We have   formally assumed that  $C<0$ (otherwise $b$ is to be rotated to
imaginary values) and picked up a solution with decreasing dilaton.

We can now generate  other  $D=4$ cosmological solutions by considering the
$O(2,2|R)$ duality
rotations  (see \gpr\ for a review)  of  (34). The simplest special  case of
the solution  (34)  with $p_1=0$ (i.e. the direct product the $x$-dependent and
$t$-dependent $D=2$ backgrounds) was discussed in
\kolu\  and its duality rotations -- in \givpas\gavem.
A  special  duality rotation of  (34) with $p_1=0$  gives the cosmological
solution of \napwi\  corresponding to the $SL(2,R) \times SU(2) /R \times R$
gauged WZW model.

It turns out that in some cases  the duality transformations  in $\t,\tt$
directions  `commute' with  the  replacement  (13)  of the moduli by functions
of time.
In particular, we  can  obtain  the same  generalisations of  (34) by
first applying the duality rotation to the spatial solution (30) and $then$
using the `rolling moduli' ansatz (13) for all the moduli (including the ones
corresponding to the parameters of the duality transformation).

For example, we can  give the following re-interpretation of the Nappi-Witten
(NW) solution
\napwi: it is
the
 cosmological generalisation of the $SU(2)$ analog of the charged
black string background \horne \ (i.e. $SU(2)\times U(1)/U(1)$  gauged WZW
model)  with the modulus
(charge)   $q$ (and the constant mode of dilaton) replaced  by  a function of
$t$.  This  solution  thus explicitly  illustrates the situation
 when a cosmological evolution   is equivalent to a
motion in the moduli space of a  `spatial' conformal field theory (equivalent
observations
were made in \kk).
	The corresponding $N=3$ spatial  background is \horne\givkir\foot {The element
of $SU(2)$ is parametrised as
follows:
$$\  g= {\rm exp}({i\over 2 } \t_L \s_2 ) {\rm exp}({{i\over 2 }
x \s_1}) {\rm exp}({{i\over 2 } \t_R \s_2  }) \ , \ \    \ \
  \  \t_L= \t + \tt \ , \ \  \t_R = \tt-\t \ .
   $$
 $q$ determines the embedding of $U(1)$ into  $SU(2)\times U(1)$.
 For a generic $q$ the metric (35)  has a
topology of $S^3$, degenerating at $q=0, -1$ (or $a_2=0, \infty$)  to  a
product of a disc and a point
(in the  $SL(2,R)$ case the corresponding limits are the $D=2$ black hole and
its dual). The $q=\infty$ limit of (35),(36)  corresponds to the $SU(2)$ WZW
model.
}
 $$  ds^2_N = dx^2 +  (1 + q)  {1-X \ov X+ 1+2q  } d\t^2
  + q {1+ X \ov X + 1+ 2q } d\tt^2 \ ,  \eq{35} $$
 $$   B_{\t \tt}   =   {q (X-1) \ov  X+1+2q}  \ ,
 \ \ \ \  \p=\p_0 - {1\ov 2} \ln (X+1 + 2q) \ ,  \ \ \   X\equiv  \cos 2b'x .
\eq{36 } $$
This background solves the leading-order Weyl invariance conditions for an
arbitrary value of $q$ (for a discussion of all-order generalisations and their
scheme dependence see \sfts).
It is dual to  (30) with
$$ a_1=1\ , \ \ \ \ a_2^2=  {q\ov q+1 }= \e{2\l}   \ , \ \ \l_1=0, \ \ \
\l_2\equiv \l \ , \
\ \ \ q=  {\e{2\l} \ov 1- \e{2\l} }\ .  \eq{37} $$
It appears that  the    $O(2,2)$ duality  transformation
`commutes' with the  replacement of  the modulus $\l$ or $q$  by a
function  of $t$:  starting with (35),(36) and replacing $q\ra q(t) $  gives a
consistent $D=4$ cosmological solution.

The action (11)
  computed  directly on the background
(35),(36) with $q\ra q(t)$ is a special case of (14),(31) (here $C$ is the same
as in (31))\foot{Note  the symmetry $q\ra -( q+1)  $  or  $a_2\ra a_2\inv$
which is the analog of the torus duality.
Its presence is obvious in view of the  relation between  (35),(36) and
 (30).}
$$ S=  \int  dt \ \e{-\vp}  \big[ C -  \dvp^2  +   {{\dot q}^2 \ov 4 q(q+1)
}\big] =
 \int  dt \ \e{-\vp}  \big( C -  \dvp^2  +   \dl^2 \big) \ .   \eq{38}  $$
The  resulting  time-dependent solution is thus
given by $ds^2_D=-dt^2 + ds^2_N (q(t)) $ with
(we choose $C<0$ and growing solution)
 $$ \l = \l_0 + \ln \tanh bt \ , \ \ \ \ \    q(t) =
{s_0^2 \tanh^2 bt \ov 1- s_0^2 \tanh^2 bt} \ ,
   \eq{39} $$
$$   4b^2= -C =   -{2(N-25)\ov 3\a' }  + 4b'^2        \ , \ \ \ \  s_0=
\e{\l_0}  \ . $$
The solution of \napwi\ is reproduced in the case when
there are no extra fields that shift the central charge,   so that for $N=3$ we
have  $C> 0$ and $b$ should be imaginary (i.e.  $\tanh $
in  (39)  should be replaced by $\tg$
so that the space has  finite life-time  $0<t<\pi/2$).

 In (39) we have $q(0) =0$ and $q(\infty) = {s_0^2 /( 1- s_0^2) } $ so
depending on $s_0$ the large $t$
limit  corresponds to   either  $S^3$ or `dual black hole'  spatial topology.
The topology change  (see  \kk\ for a detailed discussion)
thus takes place  only at the limiting points in the moduli space
which are separated by an infinite distance \cvkut.
 It
would be more  interesting to find if there exists  a cosmological realisation
of a smooth topology change taking place inside the moduli space.

Let us now consider    generalisations of
the  solution (39). As in (30) we can formally rescale $\t$ and $\tt$ in the
spatial solution (35),(36)
by two arbitrary coefficients,  replacing  the   factors  $1+q$ and $q$ by
 $(1+q)a_1^2$ and $qa_2^2 .$ We can then consider the  cosmological ansatz
(13) with  the $G_{ab},B_{ab}$
given by (35),(36) where the coefficients $a_i$ and $q$ are replaced by
functions of $t$. Such ansatz turns out to be consistent
only if
$$a_1 = a^{-1}_2 =a\equiv  \e{\l'}  \eq{40} $$
so that the determinant of the 3-metric  and $B_{\t\tt}$  retain their form
unchanged
while  the metric (35) leads to  the following $D=4$ background
 $$  ds^2_D = -dt^2 + dx^2  $$  $$
+  \ [1 + q(t)] a^2(t) {1-X \ov X+ 1+2q(t)  } d\t^2
 \  + \ q(t)a^{-2}(t) {1+ X \ov X + 1+ 2q(t) } d\tt^2 \ .  \  \eq{41} $$
The solution for  $q(t)$ and $a(t)$
is the same as in the 2-torus case (see (32),(33)), with the moduli $\l_i$
being
$\l  $ and  $\l'$ (and $\g_{ij}=\d_{ij}$).
Computing the action (11),(14) one finds   that it is given by (31)
where now
 $$ \l= \ha\  \ln \ {q\ov q+ 1} =  \l_1 +\l_2 \ , \ \ \ \   \l' =  \ln\  a =
\l_1 \ .  \eq{42}  $$
The explicit form of the solution is  found by
replacing $q$ and $a$ in (36),(41) by
$$ q=  {s_0^2 \tanh^{2(p_1+p_2) } bt \ov 1- s_0^2 \tanh^{2(p_1+p_2)} bt} \ , \
\ \ \
a^2= a_0^2 \tanh^{2p_1} bt \ , \ \ \ \   p_1^2 + p_2^2 =1 \ . \eq{43} $$
The  NW solution  \napwi\ corresponds to the special case of $p_1=0, p_2^2=1$
and
imaginary $b$.

The  background (41),(43),(36)  can be also obtained by the
$O(2,2)$ duality rotation   of the solution
(34). Namely, if (34) is written  as $$ds^2= -dt^2 + dx^2 + g_1 (x,t) d\t^2
+ g_2(x,t) d\tt^2 \ , \eq{44} $$ its duality rotation \givpas\ can be put into
the form
$$ ds^2= -dt^2 + dx^2  +  {g_1  \ov g_1g_2   + q_0^2 }\
d\t^2 +
 {g_2 \ov g_1 g_2 + q_0^2} \ d\tt^2  \ ,  \eq{45} $$ $$
B_{\t\tt }= {q_0 \ov  g_1  g_2  + q_0^2}  \ , \ \  q_0= {s_0^2 \ov 1- s_0^2} \
,
 $$
 equivalent to the solution  (41),(36).

We can also take special  limits in $q$  in (35),(36) and $then$ use rolling
moduli ansatz (13). For example,
one  can  start directly with the  $SU(2)$ WZW model background
(or $q\ra \infty$ limit of (35),(36))
 $$  ds^2 = dx^2 +    \ha {(1-X)} d\t^2
  + \ha  {(1+ X ) } d\tt^2 \ ,  \ \ \   X\equiv  \cos 2b'x \ , \eq{46} $$
 $$   B_{\t \tt}   =   \ha { (X-1) }  \ ,
 \ \ \ \  \p=\p_0  \ . \eq{47 } $$
Rescaling   $\t$  and $\tt$  one can take their `radii' and the dilaton $\p_0$
as   free parameters.
  We can also   add a constant $B_0$ to $B_{\t \tt}$.
Making these constants time-dependent one finds that    this ansatz is
consistent
only if  $a_1 = a_2\inv$ and $\dot B_0=0$  so that   $
\dot B_{\t \tt}=0 $   and at the end we   have a one-modulus problem
but with the metric in (14) $\g=2$.
As a result, the corresponding  $D=4$ cosmological metric is given by
 $$  ds^2 = -dt^2 +    dx^2   +     a_0^2  (\tanh bt )^{\sqrt 2}  \ {\rm sin}^2
b'x \ d\t^2
  + a_0^{-2}(\coth bt)^{\sqrt 2}\  {\rm cos}^2 b'x \ d\tt^2 \ .  \     \eq{48}
$$
This solution can be related to  (43) by formally taking the limit
 $q\ra \infty$, $\l\ra 0$, $\l_1\ra \l_2$ and $p_1\ra p_2 \ra \sqrt 2 $.

\def \CD {{\cal D}}

Let us now comment on the relation between the solutions  (34) (solution
`$S_1$'),
(41),(43) (solution `$S_2$') and (48) (solution `$S_3$').
Let  $\CD$ denote  a  `static'  $O(2,2)$ duality transformation
and $T$ -- the  operation of  adding a time direction to a  conformal
background and  replacing  some moduli by
functions of time in a consistent way.
 Then  acting by $\CD$ and $T$ on the  corresponding (gauged) WZW models we
have:
$$\CD (SU(2)) = [SU(2)/U(1)] \times U(1)\ ,    \  \ \ \CD (SU(2)) =
[SU(2)\times U(1)]/U(1) \ , $$
        $$     T ( [SU(2)/U(1)] \times U(1)) =    S_1\ ,  \ \ \
T  ([SU(2)\times U(1)]/U(1)) =  S_2\ , \ \ \ T  (SU(2)) =S_3\  .  $$
We have seen that $\CD (S_1$)= $S_2$, i.e. $\CD$ and $T$, in fact,
 commute when
applied to $SU(2)/U(1) \times U(1)$ model.
Note, however, that one cannot generate $S_1$ or $S_2$ by applying a duality
transformation to $S_3$ (while the first two solutions have two rolling moduli,
the third has only one).
Thus the commutativity of the duality with the rolling moduli ansatz
(13) is not universal.

The generalised time-dependent  solutions  we have
presented  above do not  have an obvious CFT interpretation: it is
only for the NW
solution \napwi\ (and its special limits and analytic continuations \kolu\kk)
that  we   know that it corresponds to a  gauged WZW model.
There is also  a similar  example  of $D=3$ cosmological solution,
which, like NW solution  is obtained from a coset model and,  at the same time,
 admits a
`rolling moduli' interpretation.
Consider the 2-dimensional part of the conformal background (30)
and rescale $x$ to introduce an additional constant (the solution of course
depends only on $b'/a_1$)
$$ ds^2_N =  a_1^2 dx^2 +  a_2^2 {\rm tan}^2 b'x d\t^2  \ , \ \eq{49} $$
$$ \ \ \p= \p_0 - \ln \cos b'x \
,  \ \ \  \tilde \beta^\p = {N-26\ov 6} -  \a'  b'^2 a_1^{-2} \ .    $$
Relaxing the $D=2$ zero  central  condition (that  fixes $b'/a_1$)
we can look for $D=3$ cosmological solutions by replacing $a_1$ and $a_2$ by
functions of $t$. Such  a solution  exists
if $a_1=a_2^{-1}$  and is given by (cf. (34))
$$  ds^2_D = -dt^2 +   a_{0}^2{\rm tanh}^{2} bt\   dx^2\  +  a^{-2}_{0} {\rm
coth}^{2} bt\  {\rm tan}^2 b'x \ d\t^2 \ , \eq{50} $$
$$
 \ \vp(x,t) = \vp_0 -
 {\rm ln \   sinh}  2 b t\ -
                          { \rm ln \ sin } 2 b'x \  .    $$
This solution is equivalent to the one obtained from the
$SO(2,2)/SO(2,1)$ gauged WZW model \bs.
This example suggests that one may be able to construct
more general  $D=4$ cosmological solutions (with (34) and (41),(43) being
special cases)
by starting  again with (30) or (35)
and introducing  three  time-dependent parameters.
Some of these models may have a coset CFT interpretation.

\newsec{Acknowledgements }
\noindent
I am  grateful to C. Schmidhuber for interesting and   stimulating discussions.
I
would like  also to acknowledge  the support of  PPARC.

\vfill\eject
\listrefs
\end

\end